\documentclass[10 pt,final,journal,letterpaper,oneside,twocolumn]{IEEEtran}
\usepackage{multicol}   
\usepackage{multirow}
\usepackage{graphicx,subfigure}
\usepackage[ruled]{algorithm2e}
\usepackage{algorithmic}
\usepackage[small]{caption}
\usepackage{float}
\usepackage{amssymb,amsmath}
\usepackage{graphicx}
\usepackage{subfigure}
\usepackage{xspace}
\usepackage{amsthm}
\usepackage{fancyhdr}

\usepackage{caption,setspace}
\usepackage{amssymb,amsmath}
\usepackage{textcomp} 
\usepackage{epstopdf}
\usepackage{bbding}
\usepackage{cite} 
\usepackage{color}
\usepackage{pifont}

\begin{document}
%
\title{Deep Reinforcement Learning for Backscatter Communications: Augmenting Intelligence in Future Internet of Things}

\author{Wali Ullah Khan, Eva Lagunas, \textit{Senior Member, IEEE,} Zain Ali, Asad Mahmood, Chandan Kumar Sheemar, Manzoor Ahmed, Symeon Chatzinotas, \textit{Fellow, IEEE,} Bj\"orn Ottersten, and \textit{Fellow, IEEE} \thanks{Wali Ullah Khan, Eva Lagunas, Asad Mahmood, Chandan Kumar Sheemar, Symeon Chatzinotas, and Bj\"orn Ottersten are with the University of Luxembourg, Luxembourg; Zain Ali is with the University of California, Santa Cruz, USA; Manzoor Ahmed is with the Hubei Engineering University, Xiaogan, China.

}}%

\markboth{Submitted to IEEE
}
{Shell \MakeLowercase{\textit{et al.}}: Bare Demo of IEEEtran.cls for IEEE Journals} 

\maketitle

\begin{abstract}
Backscatter communication (BC) technology offers sustainable solutions for next-generation Internet-of-Things (IoT) networks, where devices can transmit data by reflecting and adjusting incident radio frequency signals. In parallel to BC, deep reinforcement learning (DRL) has recently emerged as a promising tool to augment intelligence and optimize low-powered IoT devices. This article commences by elucidating the foundational principles underpinning BC systems, subsequently delving into the diverse array of DRL techniques and their respective practical implementations. Subsequently, it investigates potential domains and presents recent advancements in the realm of DRL-BC systems. A use case of RIS-aided non-orthogonal multiple access BC systems leveraging DRL is meticulously examined to highlight its potential. Lastly, this study identifies and investigates salient challenges and proffers prospective avenues for future research endeavors.
\end{abstract}

\begin{IEEEkeywords}
Backscatter communications, deep reinforcement learning, Internet-of-things.
\end{IEEEkeywords}

\IEEEpeerreviewmaketitle

\section{Introduction}
The Internet of Things (IoT) has emerged as a transformative technology, connecting a vast array of devices and enabling seamless communication and data exchange. To fully unlock the benefits of IoT networks in the next-generation era, it is crucial to establish a communication architecture that is self-sustaining and reliable for low-powered IoT devices \cite{mahmood2022comprehensive}. The future success and wide adoption of IoT networks will hinge on these devices' ability to communicate seamlessly without significantly increasing their energy consumption or incurring additional costs \cite{9857750}. As IoT networks continue to expand, there is a growing need for innovative communication technologies that can improve efficiency, reduce power consumption, and enhance connectivity. 

Various solutions have been proposed in the literature to address the challenges described above, such as device-to-device (D2D) communications, Zigbee, Rubee, Insteon, near-field communications (NFC), and Z-wave \cite{10078244}. However, none of these solutions offers the ubiquity, ease of use, and range of utility that backscatter communications (BC) provides. BC is a new paradigm that enables pervasive connectivity for low-powered wireless devices, making it particularly useful in situations where small devices need to exchange data with minimal energy consumption \cite{9551877}. 
Ambient BC, a form of BC, is gaining popularity as it allows devices to communicate by backscattering ambient RF signals. Compared to traditional wireless systems, ambient BC configurations achieve exceptional power efficiency by utilizing existing RF signals, making them ideal for large-scale deployments in various scenarios, such as industries, smart city/homes, surveillance, and entertainment \cite{wu2022survey}. 
By modulating and reflecting these signals, IoT devices can transmit data without requiring dedicated radio transmitters, thereby minimizing energy consumption and cost. Traditional BC techniques, however, often suffer from limited data rates and poor reliability \cite{10078244}. In order to surmount these limitations, researchers and engineers have turned their attention toward deep reinforcement learning (DRL), a subfield of artificial intelligence amalgamating deep learning and reinforcement learning techniques. By leveraging DRL algorithms, IoT devices can intelligently optimize their BC strategies in real time, dynamically changing environmental conditions and thus improving system performance.

The use of DRL in BC brings numerous benefits to IoT networks. First, it enables devices to dynamically adjust their transmission parameters, such as modulation schemes and transmission power, based on channel conditions, device characteristics, and quality-of-service requirements. This adaptability ensures efficient use of available resources and enhances overall network capacity. Secondly, DRL-BC can improve connectivity in challenging environments. Namely, by intelligently selecting the optimal signal reflection angles and transmission timings, IoT devices can mitigate the effects of multi-path fading, interference, and other propagation challenges, thereby both the communication reliability and range. The integration of BC-DRL also opens up new opportunities to improve the overall system performance. By considering the broader network context and leveraging DRL techniques, IoT devices can learn and adapt their communication strategies based on network-wide objectives, such as maximizing network throughput, minimizing latency, and optimizing energy efficiency (EE).

\begin{table*}[t]
\centering
\caption{Comparison of Different DRL Algorithms}
\scriptsize
\begin{tabular}{|p{2cm}| p{2cm}| p{5cm}| p{5cm}|}
\hline
Algorithm & Technique & Advantages & Disadvantages \\
\hline\hline
DQN  & Value-based & Handles high-dimensional state spaces efficiently. Provides stability and convergence guarantees. & Suffers from overestimation and can be sensitive to hyperparameters. \\ \hline 
DDQN  & Value-based & Mitigates overestimation bias of DQN. Provides more accurate value estimations. & Requires additional network updates and increased computational complexity.  \\ \hline 
DDPG  & Actor-Critic & Effective in continuous action spaces. Handles stochastic policies. & Requires careful tuning of hyperparameters. Can be sample inefficient.  \\ \hline 
PPO  & Policy-based & Stable and efficient. Handles high-dimensional state and action spaces. & Hyperparameter sensitivity. May struggle with sparse rewards.  \\ \hline 
A3C  & Actor-Critic & Utilizes parallel agents for faster learning. Works well with large action spaces. & High computational requirements. May require fine-tuning. \\ \hline 
TRPO  & Policy-based & Guarantees monotonic improvement. Handles large policy updates. & Computational inefficiency. Difficulty in handling large action spaces. \\ \hline 
\end{tabular}
\end{table*}


While current research delves into the possibilities of DRL-BS networks, there exist correlated avenues of exploration that possess substantial potential to significantly evolve the potential of such networks towards the landscape of sixth-generation (6G) and beyond. This article aims to contribute to the growing body of knowledge in this area and inspire further exploration and innovation in the realm of DRL-BC networks. Firstly, we provide a comprehensive overview of BC systems, subsequently highlighting various DRL techniques and their respective spheres of application. Then, we explore the potential areas of DRL-BC in wireless networks, followed by a comprehensive review of the existing research developments in this field. We also provide a case study on resource optimization in reconfigurable intelligent surfaces (RIS)-aided non-orthogonal multiple access (NOMA) DRL-BC network which aims showcasing the potential performance gains with such technology. Finally, we discuss current challenges and outline potential future research directions. 

\section{Overview of BC Systems and DRL Techniques}
This section studies different types of BC systems, DRL techniques, their applications, and the framework of how DRL methods could enhance BC in future IoT networks.

\subsection{Types of BC Systems}
The BC systems can be classified into two categories based on operations and carrier emitters  \cite{10078244}. 

\subsubsection{Operation-based classifications}
BC systems can be classified into two main types: passive and semi-passive. In passive BC systems, antennas utilize the energy from the incoming RF waves, while an RF source provides the necessary power for the device. Semi-passive BC systems, on the other hand, come with their own power supply, which ensures low response times and greater reliability when data is readily available. However, the use of the carrier emitter's RF signals for communication means that there is no increase in data rate.

\subsubsection{Carrier emitter-based classifications}
Based on carrier emitter, BC systems can be classified into three types, including monostatic BC, bistatic BC and ambient BC systems. In monostatic BC systems, the reader uses a single antenna to operate as a carrier emitter as well as the receiver.
In contrast, the bistatic BC systems use separate antennas for the carrier emitter and receiver. There are generally two modes of bistatic BC: bistatic co-located, where the carrier emitter and receiver are located close to each other, and bistatic dislocated BC, where they are positioned in two different locations. Ambient BC systems have the advantage of using existing RF waves for communications between the tag and receiver. This system does not require a dedicated carrier emitter but instead utilizes ambient RF signals as the carrier for data transmission.

\begin{figure*}[!t]
\centering
\includegraphics[width=0.75\textwidth]{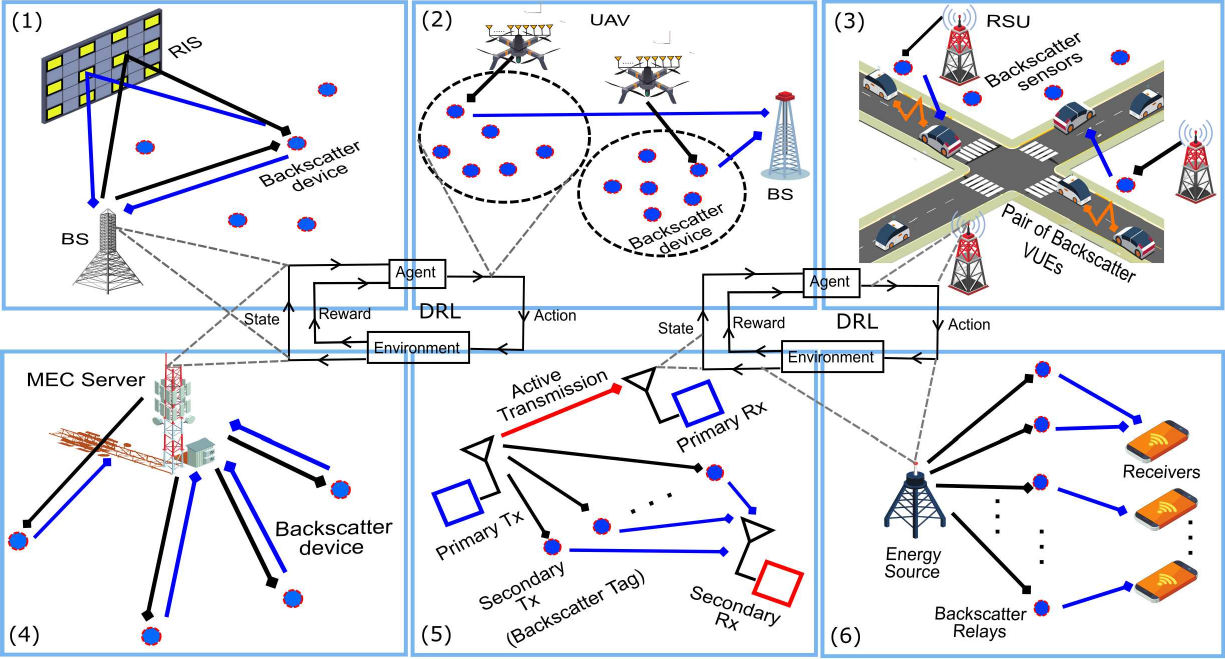}
\caption{Potential Areas of DRL-empowered BC in nextG Wireless networks. }
\label{blocky}
\end{figure*}
\subsection{Applications of DRL in BC Systems}
DRL algorithms can contribute to a diverse range of applications in BC systems. BC systems benefit from adaptive resource allocation, where DRL optimizes the utilization of limited resources such as bandwidth, capacity, and energy based on the wireless environment. DRL also enhances channel estimation by adapting estimation policies to changes in SNR and interference, leading to more accurate channel estimates and improved system performance. Moreover, DRL algorithms optimize packet scheduling, considering channel conditions, packet size, and priority to minimize collisions and interference, thereby ensuring efficient and reliable data transmission. Besides that, DRL plays a significant role in energy harvesting optimization, dynamically adjusting transmit power or transmission rate based on the current energy harvesting level, resulting in efficient utilization of harvested energy and extended device longevity. 

In addition to the above applications, DRL enhances cognitive radio capabilities by optimizing spectrum access strategies in response to changes in spectrum availability, interference, and transmission rate, maximizing the efficiency and utilization of available spectrum. DRL can intelligently choose the appropriate radio access networks based on path loss and environmental conditions, optimizing resource utilization and improving system performance. It can be beneficial also for interference management, mitigating interference in ambient BC systems through techniques such as deep Q-learning and echo state networks. Furthermore, DRL can be employed for multiple access schemes to improve the scalability and efficiency of BC systems by optimizing channel conditions for encoding and decoding. 
\subsection{Key Framework and DRL Algorithms}
DRL is a subfield of artificial intelligence and machine learning combining reinforcement learning and deep learning techniques. It involves training agents to make sequential decisions in an environment to maximize a cumulative reward signal \cite{liu2022rl}. DRL algorithms leverage deep neural networks to approximate the value or policy function, allowing for more complex and high-dimensional input spaces. In the following, we discuss some commonly used frameworks and DRL algorithms.

\subsubsection{Deep Q-Network (DQN)} DQN is a value-based DRL algorithm that uses a deep neural network to estimate the action-value function. It employs experience replay and a target network to stabilize and improve learning.

\subsubsection{Double Deep Q-Network (DDQN)} DDQN is a variant of DQN that addresses the overestimation bias in action values using two separate neural networks. It improves the accuracy of value estimation.

\subsubsection{Policy Gradient Methods} These algorithms directly optimize the policy function using gradient ascent. Examples include, Proximal Policy Optimization (PPO), and Trust Region Policy Optimization (TRPO).

\subsubsection{Actor-Critic Methods} Actor-critic algorithms combine policy-based and value-based methods. They include an actor that learns the policy and a critic that estimates the value function. Notable examples are Advantage Actor-Critic (A2C), Asynchronous Advantage Actor-Critic (A3C), and Proximal Policy Optimization (PPO).

\subsubsection{Deep Deterministic Policy Gradient (DDPG)} DDPG is an actor-critic algorithm specifically designed for continuous action spaces. It uses deep neural networks to approximate the actor (policy) and critic (value) functions and employs techniques like experience replay and target networks.

\subsubsection{Trust Region Policy Optimization (TRPO)} TRPO is a policy optimization algorithm that aims to find a policy with a significantly improved performance while ensuring that the new policy does not deviate too much from the existing policy.


\begin{table*}[tbp]
\centering
\caption{Current advances in DRL-BC systems}
\label{Rel_Works}
\scriptsize
\begin{tabular}{c c c c c} 
  	\hline
  \textbf{Ref.} & \textbf{Tag(s)}  & \textbf{Objective} &\textbf{OMA/NOMA} & \textbf{Proposed Solution} \\
  \hline \hline
  \cite{9501035} & Single  & Min BER & OMA & Joint optimization of beamforming at RIS \& reader \\
   \hline
 \cite{9248522} & Multiple & Max the EE & OMA & Designing UAV trajectory with transmit power \& reflection coefficient  \\
    \hline
\cite{9200357} & Multiple & Min the task completion time & OMA & Joint optimization of trajectory and energy  \\
	\hline
\cite{8885426}& Single & Max the system throughout & OMA & Optimization of time scheduling policy  \\
   \hline
\cite{8849964}& Multiple & Improve reward performance & OMA & Optimization of time and workload allocation  \\
   \hline
\cite{10110962}& Multiple & Max the system throughput & OMA & Optimization of mode selection and clustering  \\
   \hline
\cite{9058982}& Multiple & Max the system sum rate & OMA & Optimization of user scheduling  \\
   \hline
\cite{9197625}& Two & Max the system throughput & OMA & Optimization of reflection coefficients  \\
   \hline
[Our]& Multiple & Max the EE & NOMA & Optimization of transmit power, tag selection, reflection coefficients, and phase shift design  \\
   \hline
\end{tabular} 
\end{table*}

\section{Potential Areas and Recent Advances}

DRL-BC is a rapidly advancing field with several potential areas in IoT networks. In the following, we will discuss these areas in detail and report on recent advances in DRL-BC.

\subsection{Potential Areas}
As shown in Figure \ref{blocky}, DRL-BC has the potential to enhance the performance of various communication areas, such as RIS-aided cellular network, unmanned aerial vehicles (UAVs), Internet of vehicles (IoV), mobile edge computing (MEC), cognitive radio (CR), and relay networks. In the following, we discuss these areas in detail.
\subsubsection{DRL-BC in RIS Networks}
RIS is crucial in enhancing wireless communications performance by intelligently manipulating signal direction. By configuring the reflection properties of the surface, RIS improves signal strength, expands coverage, mitigates interference, and enhances communication quality. BC systems faced a challenge from direct-link interference caused by unknown ambient signals, resulting in lower detection performance and reduced reliability. Integrating DRL-BC with RIS offers a solution to enhance system performance regarding EE, extended coverage, and improved reliability. 
\subsubsection{DRL-BC in UAV Networks}
A conventional BC system with a fixed-located carrier emitter and receiver faces a double channel-attenuation challenge. Moreover, the BC system with co-located RF source and reader suffers from severe self-interference, which limits its transmission distance. Further, BC systems also experience fairness issues in transmission rates and energy outage problems due to the distance between carrier emitter, tag and receiver. Benefiting from the high mobility and flexible deployment, UAVs can provide emergency coverage and shorten the distance between the tag and the receiver. When no terrestrial infrastructure is available to provide ambient RF signals for backscattering, UAVs can provide signals to enable BC.
\subsubsection{DRL-BC in IoV Networks}
DRL-BC is an effective IoV solution for efficient vehicle-to-vehicle and vehicle-to-infrastructure communications. BC provides effective communications between vehicles and BS by using the processing power of DRL algorithms. Due to power consumption and enhanced communications range, BC significantly improves the dependability of IoV networks. Moreover, BC provides a low-cost option because of its low hardware requirements, making it a good candidate for IoV's wide-scale implementation. Combining DRL-BC with IoV can improve traffic management, safety, and productivity.

\subsubsection{DRL-BC in MEC Networks}
DRL-BC in MEC can revolutionize data processing and storage by bringing it closer to the network's edge, enabling a wide range of IoT applications such as augmented reality, video analytics, and edge computing. By harnessing the power of DRL algorithms, BC within the MEC framework offers low-power and reliable communication for IoT and edge devices. This integration empowers real-time processing, facilitates rapid response times, and enables seamless data transmission between IoT devices and MEC servers. The synergy between DRL-BC and MEC unlocks the full potential of IoT services, ensuring improved performance, enhanced connectivity, and increased efficiency at the network's edge. 
\subsubsection{DRL-BC in CR Networks}
Integrating DRL-BC with CR networks represents a transformative leap in wireless communications. Combining DRL's autonomous learning capabilities with the intelligent spectrum management of CR networks enables backscatter devices to dynamically adapt their modulation and reflection processes based on real-time environmental feedback. It empowers efficient and reliable communication for low-power backscatter devices, revolutionizing applications such as smart cities, IoT deployments, and industrial automation, where seamless connectivity and autonomous operation are paramount.
\subsubsection{DRL-BC in Relay Networks}When combined with relay networks, BC emerges as a promising technology, providing EE communications between transmitters and receivers. The energy-saving advantages for the network are made possible by the low-complexity design of BC, which allows relay nodes to conserve energy while being inactive. This integration unlocks extended communication range, improved signal, and EE, thus opening up new possibilities in the IoT, smart cities, and remote sensing systems. 
\subsection{Recent Advances in DRL-BC Systems}
This section discusses the recent developments in the realm of DRL-BC, by highlighting the prominent contributions from the DRL-BC research community \cite{9501035,9248522,9200357,8885426,8849964,10110962,9058982}. Jia {\em et al.} \cite{9501035} have investigated bit error rate (BER) in RIS-aided BC systems. They jointly optimized the passive beamforming of RIS and reader using the DRL technique without channel information. The authors in \cite{9248522} have maximized the EE of UAV BC systems by designing the trajectory of the UAV while considering the reflection coefficient, transmit power, and fairness. Zhang {\em et al.} \cite{9200357} have minimized the task completion time in the UAV BC system by jointly optimizing the trajectory and energy of the system. Moreover, the researchers in \cite{8885426} have optimized the time scheduling policy in cognitive radio BC to maximise the system throughput. Xie {\em et al.} \cite{8849964} have optimized the BC system's time and workload allocation to improve reward performance, stability, and learning speed.
Furthermore, the work in \cite{10110962} has maximized the system throughput in BC systems by optimizing the mode selection and clustering. Similarly, the research work in \cite{9058982} has optimized the user scheduling in symbiotic radio-aided IoT networks to maximize the sum rate. Table II shows a detailed comparison of current advances in DRL-BC systems. 

\section{Case Study: Resource Optimization for RIS-aided NOMA BC Systems using DRL}
This section provides a new optimization framework to maximize the EE and spectral efficiency (SE) of the RIS-aided NOMA BC system using the DRL technique. It will provide a system model, discuss problem formulation and proposed DRL solution, followed by numerical results.
\subsection{System Model and Proposed Solution}
\begin{figure}[!t]
\centering
\includegraphics[width=0.40\textwidth]{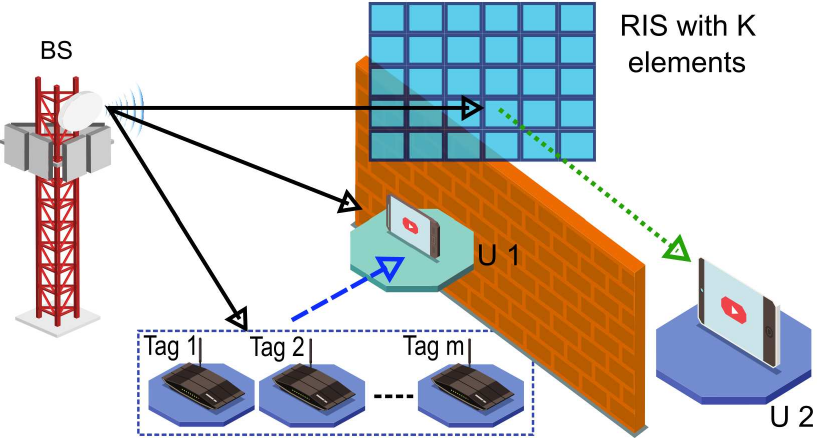}
\caption{System model}
\label{SM}
\end{figure}

As illustrated in Fig. \ref{SM}, we consider a RIS-aided NOMA BC system that consists of one BS, $M$ backscatter tags, one RIS with $K$ elements, and two NOMA users, i.e., $U_1$ and $U_2$. More specifically, $U_1$ receives signals from BS and tags, while $U_2$ is blocked by obstacles and incapable of receiving from BS through the direct link. To address this issue, we consider RIS to assist the BS signal to $U_2$ \cite{10003076}.
Following the downlink NOMA communication, BS sends a superimposed signal to $U_1$ through a direct link and $U_2$ through a RIS-aided link. Due to the broadcast nature, tags also receive signals from BS. Tags harvest energy from the power of the received signal, using a portion of it for circuit operations and remaining for backscattering. Thus, $U_1$ receives the transmitted signal from BS and the backscattering signal from tags. This makes $U_1$ a NOMA user and a backscatter receiver. 

We seek to maximize the system EE and SE by optimizing the transmit power of BS, tag selection, and phase shift design at RIS. To achieve these objectives, we formulate non-convex optimization problems for EE and SE. For practical reasons, we consider imperfect successive interference cancellation (SIC), where the user still faces some interference even after performing SIC. The considered problems are also subjected to several practical constraints, such as maximum transmit power constraint at BS, NOMA power allocation constraint, tag selection, minimum required rate constraint, and phase shift design constraint. It might be possible to solve them using conventional optimization frameworks. However, conventional iteration-based optimization frameworks are very slow, and convergence is not guaranteed in the case of non-convex problems. Moreover, the fractional objective (in the case of EE) and non-convexity of the constraints are very complex and make the problem challenging to solve using a conventional optimization framework. Therefore, we adopt DRL frameworks which provide excellent performance for such complicated problems.
\begin{figure*}
\centering     
\subfigure[]{\label{fig:first}\includegraphics[width=58mm]{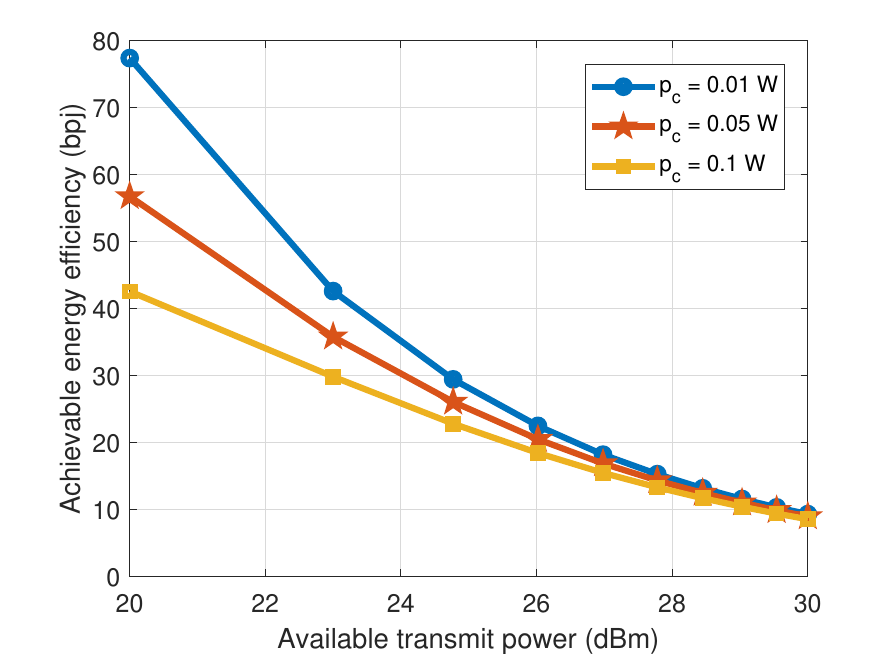}}
\subfigure[]{\label{fig:second}\includegraphics[width=58mm]{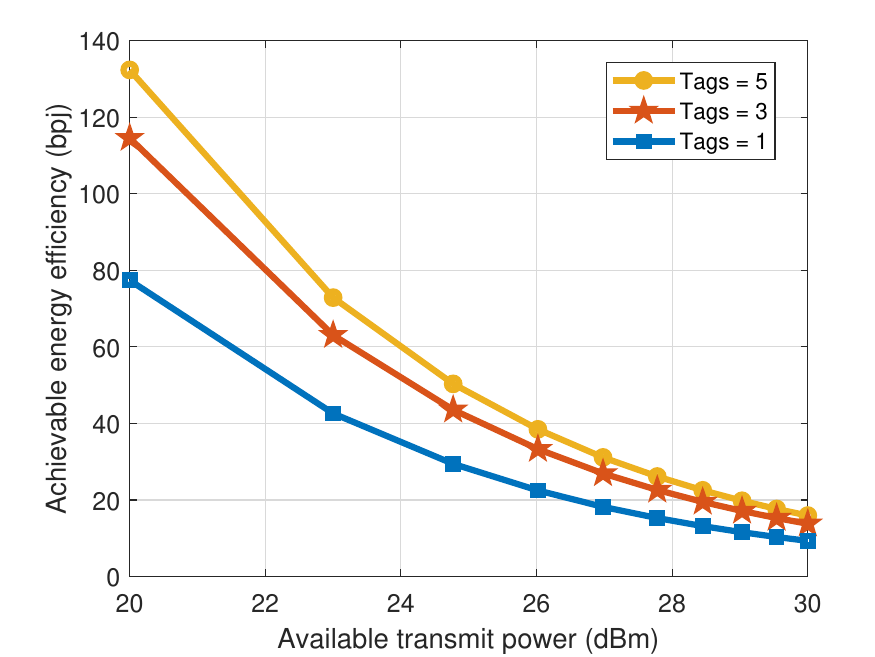}}
\subfigure[]{\label{fig:third}\includegraphics[width=58mm]{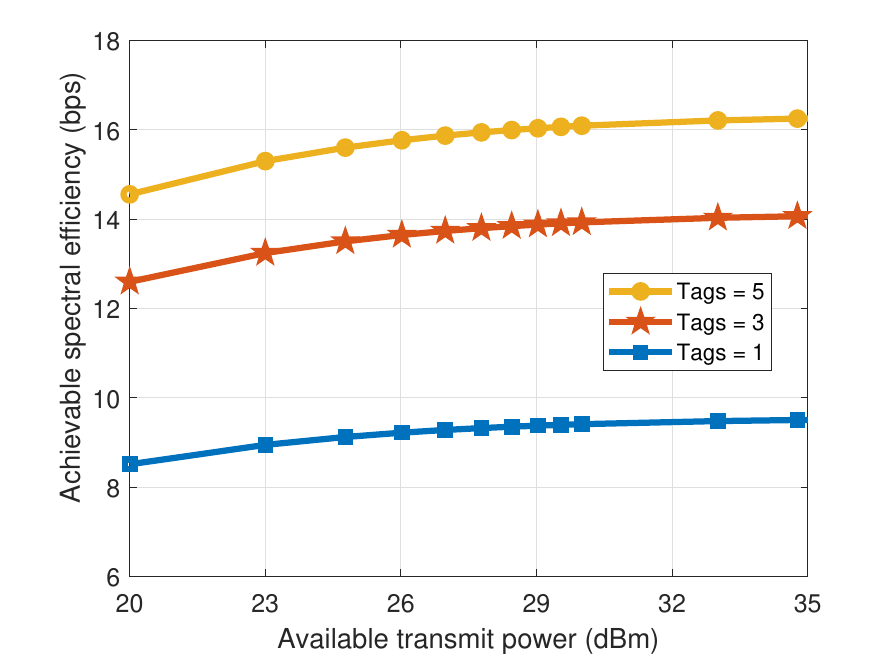}}
\caption{Impact of increasing transmission power on the system's EE and SE for different values of $p_c$ and tags.}
\label{fig:figures}
\end{figure*}

For our solution, we employ DRL-based frameworks. DRL frameworks differ from supervised learning approaches because they do not require pre-existing training data. Instead, DRL agents learn optimal actions by interacting with the environment and receiving feedback accordingly. This characteristic makes DRL a robust learning scheme, as it effectively adapts to dynamic and changing environments. By actively engaging with the environment, DRL enables our solution to handle varying conditions with exceptional flexibility and adaptability. We employed two distinct DRL agents. Both agents utilize Q-function values to learn the optimal policy. The first DRL agent focuses on optimizing the phases of the IRS elements. Its deep neural network architecture consists of two hidden layers: the first layer comprises 1000 neurons, while the second layer contains 500 neurons. The input to this agent is the channel gains from the BS to the RIS and from the RIS to user. The output of this agent is the phase values for the RIS elements. On the other hand, the second DRL agent handles power allocation values and backscatter optimization. Its deep neural network also has two hidden layers, with the first layer having 200 neurons and the second having 100 neurons. The input for this agent includes all the channel gains (where the overall channel gains of RIS are considered, including the phase results of the first DRL) in the system, and the output consists of the power allocation values and the decision regarding tag selection. Both neural networks employ Rectified Linear Unit (ReLU) activation functions. 

We opted to train two separate DRL agents instead of a single one due to the independence of optimal phases of the RIS elements from the power allocation values (the optimal power allocation values depend on the phases of RIS, but the optimal phases of RIS is not impacted by the power allocation values). By separating the problem into two independent sub-problems and training individual DRL agents, we reduce training complexity and obtain better and faster solutions. The loss function for both DRL agents is the mean square error, and we utilize the Adam optimizer to optimize the weights and biases of the neural networks. Our framework benefits from memory replay, where we maintain a memory buffer of size 50,000 samples. During each training iteration, the agents randomly sample the memory buffer and select a minibatch of 1000 samples for training purposes. As demonstrated in the subsequent subsection, the hyper-parameter values chosen for our framework have proven to deliver excellent performance for the specific problems under consideration.

\subsection{Results and Discussion}
This section provides simulation results of the considered problem, which is solved with DRL. For simulation, the value of circuit power consumption ($p_c$) was 0.01 W, the noise variance was 0.1, the RIS had 25 elements, and the value of the imperfect SIC factor was 0.2. The DRL agents were trained for 200,000 independent samples, and the testing was performed for 20,000 samples.

Fig. \ref{fig:first} illustrates the relationship between increasing transmission power and its impact on EE, considering different $p_c$ values at the BS. As transmission power increases, EE decreases due to the logarithmic nature of the rate function. The diminishing marginal gains in the rate compared to the increase in transmission power result in reduced EE. The figure also reveals that lower $p_c$ values correspond to improved EE within the system. This observation emphasizes the significance of hardware efficiency in achieving optimal energy utilization. Efficient hardware components play a crucial role in minimizing power losses and maximizing EE, thereby contributing to overall performance.

Increasing the number of backscatter tags in the system enhances EE as shown in Fig. \ref{fig:second}. With a larger pool of backscatter tags to choose from, the optimization framework can leverage the diversity in channel conditions, ensuring reliable and efficient data transmission to the user. This increased flexibility allows the system to adapt to varying environmental factors, enhancing energy consumption while maintaining the same power consumption at the BS.

Fig. \ref{fig:third} shows the impact of enhanced transmission on the SE of the system for different numbers of tags. As the transmission power increases, the SE also increases. However, the rate of increase in the SE diminishes with higher transmission power due to the logarithmic nature of the rate function mentioned earlier. Additionally, a higher number of backscatter tags leads to an increased SE, as it introduces more favorable channel options for data transmission.

\section{Current Challenges and Future Directions}
This section highlights some key challenges and outlines some potential future research directions.
\subsection{Current Challenges}

\subsubsection{Hardware Implementation}
There has been minimal research done on hardware implementations of learning-based approaches for the physical layer in BC systems due to their novelty. At present, most DRL-BC techniques are still confined to simulation stages and require significant improvement before they can be practically implemented. To enhance these models, improvements in communication channels are necessary, as the channels used for DRL simulations are primarily based on models. Real-time measurements of BC channels can help significantly in this regard.

\subsubsection{Scalability of BC Systems}
Scaling DRL-BC is a significant challenge. The auto-encoders approach is less effective when the system complexity increases—moreover, basic end-to-end communications experience exponential complexity with a growing network. Instead of completely transitioning to DRL, it is better to augment DRL for specific tasks related to BC. Another exciting approach is the deep unfolding of existing communication schemes, which can improve signal processing techniques. To ensure the scalability of these models, future research on BC should focus on reducing training and model complexity, given the rich datasets that will be available.

\subsubsection{Appropriate Model Selection}
Choosing the suitable DRL model is crucial for BC because a single model cannot fit all situations. Supervised learning techniques rely heavily on labeled data, which cannot optimize unknown parameters of BC. On the other hand, unsupervised learning techniques do not require labeled parameters and are suitable for behavior classification and fault detection in communication. DRL optimize performance through iterative decision-making processes. Due to the diversity of DRL techniques, selecting the appropriate model for BC is highly important. Future research should focus on developing a comprehensive set of DRL models that can be used to optimize different aspects of BC systems.

\subsubsection{Massive Data}
Although DRL models often outperform conventional optimization approaches, they require a substantial amount of high-quality data, especially when training for large and complex communication architectures. However, data collected by backscatter devices in practical conditions may suffer from class imbalance, loss, and redundancy, rendering it unsuitable for direct use in training and testing. Therefore, to implement intelligent BC network architecture effectively, a significant
amount of high-quality data and mature streamlining platforms are required. Industry stakeholders and researchers can address this by releasing backscatter datasets and measurements that can be used to train and test DRL models.

\subsection{Future Research Directions}
\subsubsection{Low-power IoT Networks}
BC can solve the energy consumption issues of low-power IoT devices. These devices can harvest energy from neighboring Wi-Fi networks or other wireless devices without an external power supply. BC faces challenges due to low signal-to-noise ratio and transmission range. DRL can improve BC performance by optimizing modulation, reducing interference, and improving signal quality. DRL algorithms can help IoT devices transmit data more efficiently by finding the best modulation scheme.
\subsubsection{Smart cities}
BC regulates many urban infrastructure, including traffic signals, lighting, and parking meters. This technology could improve public safety, EE, and transportation. BC in smart city applications involves scalability, security, and flexible traffic and weather response. DRL approaches can increase functionality by predicting traffic congestion, improving EE, and detecting urban abnormalities. DRL can predict traffic patterns and change traffic signal timing to alleviate congestion and improve road safety.

\subsubsection{Healthcare monitoring Systems}
BC lets people remotely check vital signs and health indicators without wearing bulky devices or being attached. This novel strategy may improve patient care and lower healthcare expenses. However, BC in healthcare applications requires precision and reliability. DRL reduces noise, improves signal quality, and discards extraneous information to improve these systems. DRL can monitor a patient's vital signs and alert clinicians to any irregularities before they become life-threatening.

\subsubsection{Environmental monitoring Systems}
BC can monitor temperature, humidity, and air quality even in remote areas. This system could track climate change, pollutant hotspots, and natural calamities. BC must resist interference, harsh weather, and other environmental challenges in environmental applications. DRL may learn to recognize and filter noise, optimize sensor parameters, and detect environmental changes over time to increase these systems' accuracy and dependability.

\section{Conclusion}
BC is an emerging solution for IoT applications in the next-generation communications era. The utilization of DRL techniques can significantly improve its performance. This article has extensively reviewed the latest research on DRL for BC and identified its key applications. It has also explored the potential areas of DRL-BC in wireless networks. A new optimization framework to maximize the spectral and EE of RIS-aided NOMA BC system has also investigated using DRL technique. Finally, the article highlights the existing challenges and future research directions in BC through a detailed gap analysis.

\ifCLASSOPTIONcaptionsoff
  \newpage
\fi

\bibliographystyle{IEEEtran}
\bibliography{Wali_EE}

\section*{Biographies}\small
\noindent {\bf Wali Ullah Khan [M]} (waliullah.khan@uni.lu) received a Ph.D.
degree in information and communication engineering from
Shandong University, China, in 2020. He is currently
working with the SIGCOM Research Group, SnT, University of Luxembourg.
\vspace{0.1cm}

\noindent {\bf Eva Lagunas [SM]} (eva.lagunas@uni.lu) received a Ph.D. degree
in telecommunications engineering from the Polytechnic University of Catalonia (UPC), Barcelona, Spain, in 2014. She currently holds a research scientist position in the SIGCOM Research
Group, SnT, University of Luxembourg.
\vspace{0.1cm}

\noindent {\bf Zain Ali} (zainalihanan1@gmail.com) received his Ph.D. degree
in electrical engineering from COMSATS University, Islamabad,
Pakistan, in 2021. Currently, he is working as a postdoctoral researcher in the Department of Electrical and
Computer Engineering, University of California, Santa Cruz.
\vspace{0.1cm}

\noindent {\bf Asad Mahmood [S]} (asad.mahmood@uni.lu) received
his Master degrees in Electrical Engineering from COMSATS University Islamabad, Wah Campus, Pakistan. He is currently pursuing the Ph.D. degree, in the SIGCOM Research Group, SnT, University of Luxembourg.
\vspace{0.1cm}

\noindent {\bf Chandan Kumar Sheemar} (chandankumar.sheemar@uni.lu) received his Ph.D. degree from EURECOM, Sophia Antipolis, France, in 2022. He is currently a research associate in the SIGCOM Research Group, in the SIGCOM Research Group, SnT, University of Luxembourg, Luxembourg. 
\vspace{0.1cm}

\noindent {\bf Manzoor Ahmed} (manzoor.achakzai@gmail.com) received a
Ph.D. from Beijing University of Posts and Telecommunications China in 2015. He is currently a professor with the School of Computer and Information Science and also with the Institute
for AI Industrial Technology Research, Hubei Engineering University, Xiaogan, China.
\vspace{0.1cm}

\noindent {\bf Symeon Chatzinotas [F]} (symeon.chatzinotas@uni.lu) received
Ph.D. degrees in electronic engineering from the University of Surrey, Guildford, United Kingdom, in 2009. He is currently a full professor or Chief Scientist I and the co-head of the SIGCOM Research Group, SnT, University of Luxembourg.
\vspace{0.1cm}

\noindent {\bf Bj\"orn Ottersten [F]} (bjorn.ottersten@uni.lu) received his Ph.D.
degree in electrical engineering from Stanford University, California, in 1990. He is currently the director for SnT, University of Luxembourg.
\end{document}